\documentclass[aps,pra,superscriptaddress,twocolumn,floatfix,longbibliography]{revtex4-1}
\usepackage{dcolumn}
\usepackage{graphicx,color}
\usepackage[utf8]{inputenc}
\usepackage{amsmath,amssymb,bm,amsfonts,dsfont,mathrsfs,amsthm}
\usepackage{braket}
\usepackage{txfonts,comment}
\usepackage{url}
\usepackage[colorlinks=true,linkcolor=blue,citecolor=blue,urlcolor=blue]{hyperref}
\usepackage{soul}

\begin{document}

\title{Cavity-Controlled Hysteresis and Multilevel Molecular Switching in Hybrid Optomechanics}

\author{Hugo Molinares}
\email{molinareshugo@gmail.com}
\affiliation{Facultad de Ciencias, Universidad San Sebastián, Ciudad Universitaria, Avenida del Cóndor 720, 8580704, Huechuraba, Santiago, Chile}

\author{Raúl Coto}
\affiliation{Department of Chemistry and Physics, Nova Southeastern University, Fort Lauderdale-Davie, Florida 33328-2004, USA}

\author{Vitalie Eremeev}
\affiliation{Centro Multidisciplinario de Física, Vicerrectoría de Investigación, Universidad Mayor, 8580745 Santiago, Chile}

\author{Eduardo Chamorro}
\affiliation{Facultad de Ciencias, Universidad San Sebastián, Ciudad Universitaria, Avenida del Cóndor 720, 8580704, Huechuraba, Santiago, Chile}

\begin{abstract}
We investigate cavity-controlled hysteresis and multilevel molecular switching in a hybrid molecular optomechanical system composed of a plasmonic nanocavity, a Fabry-P\'erot cavity, and an ensemble of molecular vibrations. Molecular-plasmonic optomechanical interaction together with cavity-plasmon hybridization produce bistability and dynamical hysteresis, leading to a history-dependent optical and molecular response. An analytical treatment of the steady-state equations shows that the auxiliary cavity renormalizes the effective optical linewidth and detuning, while the collective molecular interaction generates an effective nonlinearity. Together, these effects enable control of the hysteretic regime through optical detuning, cavity-plasmon coupling, and molecular ensemble size. Beyond bistability, pulsed excitation generates staircase-like molecular dynamics, with ascending or descending vibrational populations depending on the optical driving configuration. In particular, the FP cavity controls the buildup of vibrational excitation, while simultaneous driving of both optical modes enables stepwise vibrational depletion. These results establish hybrid molecular optomechanical systems as a tunable platform for history-dependent and pulse-driven molecular dynamics, with potential applications in nanoscale memory, multilevel information processing, and optically controlled molecular switching.
\end{abstract}

\maketitle

\section{Introduction}

Molecular cavity optomechanics provides a nanoscale platform in which high-frequency molecular vibrations interact with strongly confined optical fields. Owing to the small effective mode volumes attainable in plasmonic nanocavities, molecular vibrations can exhibit comparatively large optomechanical coupling strengths while their high resonance frequencies suppress thermal occupation at room temperature \cite{Roelli2016,Esteban2022}. These properties have motivated a broad range of applications, including surface-enhanced and coherent Raman scattering \cite{Roelli2016,Esteban2022,Tabatabaei2015,HuangLeiZhang2025}, optomechanically induced transparency \cite{Yin2025}, frequency conversion and infrared detection \cite{Xomalis2021,Zou2024}, heat transfer \cite{Ashrafi2019}, precision sensing \cite{Liu2018a}, collective molecular effects \cite{Jakob2023,Huang2024}, quantum heat engines \cite{Zhu2025}, collective entanglement \cite{Huang2024,Yu2026}, phonon lasing \cite{Mol_phonon_laser}, and photon blockade \cite{Mol_blockade}.

In addition, Fabry-P\'erot (FP) cavities offer versatile means of controlling optomechanical interactions through tailored driving schemes, optical detuning, and dissipation engineering. Such approaches have been exploited to achieve dynamical backaction cooling \cite{Marquardt2007}, mechanical amplification and self-oscillation \cite{Kippenberg2005}, phonon lasing \cite{Grudinin2010,Eremeev2025}, mechanical squeezing \cite{Kronwald2013}, and the generation and control of quantum correlations \cite{Montenegro2019,Vitali2007,Molinares2025}. Multimode configurations and tailored driving further enhance control over optomechanical dynamics \cite{Kuzyk2017,Wang2012}.

In this context, a significant advance is the integration of plasmonic nanocavities with high-quality photonic resonators. Hybrid plasmonic-photonic architectures combine strong field confinement with the spectral selectivity and tunability of dielectric or FP cavities \cite{acsnano.4c16535,Shlesinger2023,Roelli_Review2024,mean_field_mol_opt,Dezfouli2019,Barreda2021}. Experimentally, hybrid cavity-antenna systems have demonstrated tunable optical resonances and strong Raman enhancement, providing a route toward dynamical backaction in molecular optomechanics \cite{Shlesinger2023}. Related experiments in molecule-filled plasmonic nanocavities have reported hybrid properties arising from molecular optomechanical interactions \cite{PatraKafleHabteyes2023}. Theoretical studies have explored hybrid molecular optomechanical platforms for optical transmission and slow-fast-light control \cite{Yin2025}, collective quantum correlations \cite{Huang2024,Yu2026}, and nonclassical photon statistics \cite{Mol_blockade}.

The nonlinear dynamics of hybrid molecular-plasmon-FP cavity architectures remain relatively unexplored. Optomechanical systems can exhibit bistability and hysteresis under strong driving \cite{Deeg2025, Ariel2026}, while molecular vibrations introduce additional nonlinearities through intensity-dependent backaction. Hysteretic switching has been explored in molecular optomechanics \cite{Zhu2025}, and pulsed excitation can generate memory effects and discrete phononic transitions \cite{Ariel2026}. Intercavity coupling can also modify optical bistability \cite{She2025}. Other forms of memory have also been studied in cavity-embedded molecules, including memory-assisted coherence \cite{Kinslow2026}. However, controlling these phenomena through cavity-plasmon hybridization in molecular systems remains largely unexplored, particularly the tunability of bistability and pulse-driven collective vibrational dynamics.

In this work, we address these questions in a hybrid molecular optomechanical system composed of a plasmonic nanocavity, a FP cavity, and an ensemble of molecular vibrations. We show that the FP cavity provides a direct control mechanism for the nonlinear molecular-optomechanical response. First, we demonstrate simultaneous dynamical hysteresis in the plasmonic cavity, FP cavity, and collective molecular-vibrational populations and derive analytical steady-state equations that identify how the cavity-plasmon coupling renormalizes both the effective optical linewidth and detuning, while the collective molecular interaction generates an effective nonlinearity. This description allows us to determine the bistability condition and to show how the bistable region can be controlled through the cavity–plasmon coupling strength, optical detuning, and the number of molecules $N$. Second, we extend the analysis to the pulse-driven dynamics of the collective molecular vibrational mode and demonstrate the emergence of staircase-like excitation levels. The auxiliary FP cavity provides an additional control channel, allowing higher vibrational populations to be reached at earlier times and enabling ascending or descending staircases depending on the optical driving configuration. Thus, our results connect cavity-controlled bistability with pulse-driven collective molecular switching within a single hybrid optomechanical architecture.

The paper is organized as follows. Sections \ref{sec2} and \ref{sec3} introduce the hybrid molecular optomechanical system and its mean-field description. Section \ref{sec4} examines dynamical hysteresis, steady-state bistability, and parameter control. Section \ref{sec5} explores pulse-driven molecular excitations and their control through cavity-plasmon coupling. Section \ref{sec6} discusses tunability and experimental feasibility. Finally, Sec. \ref{sec7} presents the conclusions.

\section{Model}\label{sec2}

\begin{figure}[t]
\centering
\includegraphics[width=0.8\linewidth]{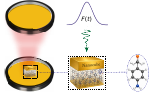}
\caption{Schematic of the hybrid molecular optomechanical system. An Au nanoblock deposited on one of the mirrors forms a NPoM plasmonic nanocavity with resonance frequency $\omega_a$, driven by the time-dependent field $F(t)$. A molecular ensemble with vibrational frequency $\omega_m$ is located within the nanocavity hotspot and interacts with the plasmonic field. The two Au mirrors form a FP cavity supporting an optical mode with resonance frequency $\omega_c$.}
\label{fig1}
\end{figure}

As illustrated in Fig. \ref{fig1}, we consider a hybrid molecular cavity-optomechanical (COM) system composed of an Au nanoblock, a molecular ensemble, and two highly reflective Au mirrors. The plasmonic mode is driven by a time-dependent field of amplitude $F(t)$ and frequency $\omega_l$. The Au nanoblock is positioned on one of the mirrors, forming a nanoparticle-on-mirror (NPoM) configuration \cite{npom2022,npom_adm} that supports a localized plasmonic mode with resonance frequency $\omega_a$.

Together, the two Au mirrors form a FP cavity supporting an optical mode with resonance frequency $\omega_c$. The plasmonic and FP cavity modes are coupled through their evanescent fields with coupling strength $J$ \cite{Shlesinger2023,mean_field_mol_opt}. The molecular ensemble is confined within the plasmonic hotspot, where molecular vibrations of resonance frequency $\omega_m$ interact with the localized plasmonic field through the single-molecule optomechanical coupling $g$ \cite{opt_coup,Roelli2016}. Owing to the strong field confinement of the NPoM geometry, this interaction is significantly enhanced \cite{opt_coup,Jakob2023}. The corresponding Hamiltonian can then be written as ($\hbar=1$)
\begin{eqnarray}
H &=& \omega_a a^\dagger a
+ \omega_c c^\dagger c
+ \sum_{i=1}^{N}\omega_m b_i^\dagger b_i +  J\left(a^\dagger c + ac^\dagger\right)\nonumber\\
&-& \sum_{i=1}^{N}g\,a^\dagger a
\left(b_i^\dagger+b_i\right)+ iF(t)
\left(a^\dagger e^{-i\omega_l t}
-ae^{i\omega_l t}\right),
\label{eq:H1}
\end{eqnarray}
\begin{figure*}[t]
\centering
\includegraphics[width=0.8\linewidth]{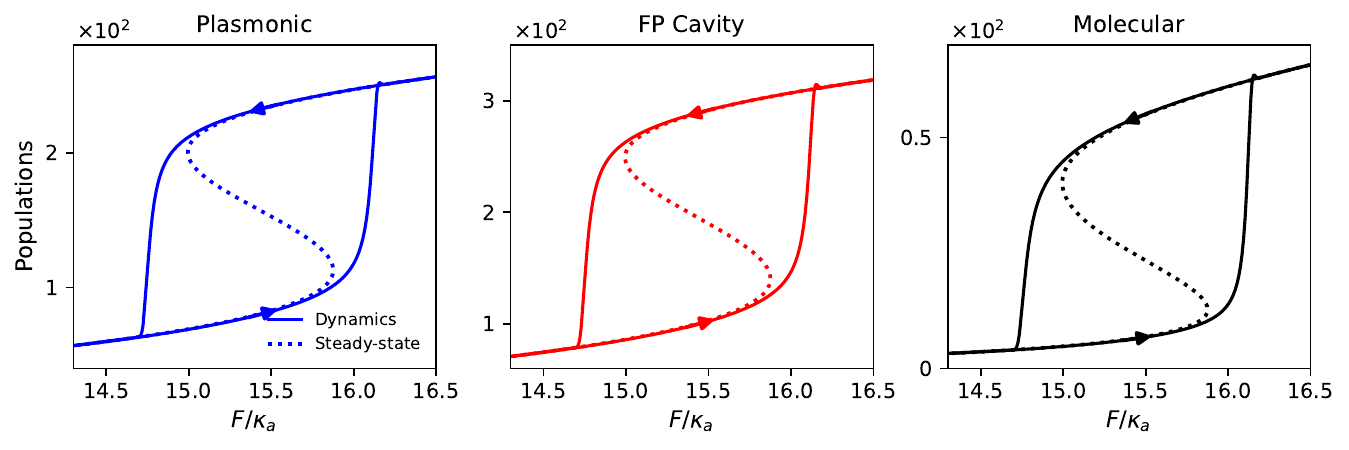}
\caption{Dynamical hysteresis in the plasmonic cavity, FP cavity, and molecular-vibrational populations under a triangular modulation of the driving amplitude $F$. Solid curves correspond to the time-dependent mean-field dynamics during the forward and backward sweeps, whereas dotted curves show the associated steady-state branches. The parameters (in units of $\kappa_a$) are: $\omega_m=4.5$, $\Delta_a=\Delta_c=-8.7$, $g=0.015$, $J=9.7$, $\kappa_c=10^{-2}$, $\gamma_m=10^{-3}$, $\Delta F=25$, $t_s=5\times10^3$, and $N=90$.}
\label{fig2}
\end{figure*}
where $a$ ($a^\dagger$), $c$ ($c^\dagger$), and $b_i$ ($b_i^\dagger$) are the annihilation (creation) operators for the plasmonic nanocavity optical mode, FP cavity optical mode, and $i$-th individual molecular
vibrational mode, respectively. $F(t)$ is the time-dependent driving amplitude. Here, we consider a large molecular ensemble, $N\gg1$, and investigate how the molecular number $N$ influences the hysteretic response of the hybrid system. To this end, we introduce the collective molecular-vibrational mode $\hat{B}=\frac{1}{\sqrt{N}}\sum_{i=1}^{N}\hat{b}_i$ \cite{Huang2024,Zou2024}, which captures the collective dynamics of the molecular ensemble. In the rotating frame at the driving frequency $\omega_l$, the system Hamiltonian can then be written in the simplified form
\begin{eqnarray}
H &=& \Delta_a a^\dagger a +\Delta_c c^\dagger c
+\omega_m B^\dagger B + J\left(a^\dagger c + ac^\dagger\right)\nonumber\\
&& 
- g_N a^\dagger a\left(B^\dagger+B\right) +iF(t)\left(a^\dagger-a\right),
\label{eq:H}
\end{eqnarray}
where $\Delta_{a,c}=\omega_{a,c}-\omega_l$, and $g_N=g\sqrt{N}$ is the optomechanical coupling enhanced by the collective effect of the molecules \cite{Huang2024, Zou2024}. Including dissipation and the associated input-noise operators, the dynamics of the system is described by the Heisenberg-Langevin equations
\begin{align}\label{lang_eqs1}
\dot{a} &= \left(-i\Delta_a-\kappa_a\right)a-iJc
+ig_Na\left(B+B^\dagger\right)
\nonumber\\
&+F(t)+\sqrt{2\kappa_a}\,a_{\mathrm{in}}(t),
\\
\dot{c} &= \left(-i\Delta_c-\kappa_c\right)c-iJa
+\sqrt{2\kappa_c}\,c_{\mathrm{in}}(t),
\\
\dot{B} &= \left(-i\omega_m-\gamma_m\right)B
+ig_Na^\dagger a
+\sqrt{2\gamma_m}\,B_{\mathrm{in}}(t),
\label{lang_eqs2}
\end{align}
where $\kappa_a$, $\kappa_c$, and $\gamma_m$ are the decay rates of the plasmonic, FP cavity, and collective molecular-vibrational modes, respectively. The operators $a_{\mathrm{in}}(t)$, $c_{\mathrm{in}}(t)$, and $B_{\mathrm{in}}(t)$ describe the corresponding reservoir fluctuations. Assuming zero-mean input reservoirs, as appropriate for vacuum and thermal bath states, the input-noise operators satisfy the conditions \cite{Aspelmeyer2014, Gardiner1985}, $\langle a_{\mathrm{in}}(t)\rangle=
\langle c_{\mathrm{in}}(t)\rangle=
\langle B_{\mathrm{in}}(t)\rangle=0$.
Consequently, these noise terms do not contribute explicitly to the equations governing the average amplitudes. 

To investigate dynamical hysteresis, we consider a triangular modulation of the drive amplitude applied to the plasmonic nanocavity, following the protocol employed in Ref. \cite{Casteels2016}. Specifically, the drive amplitude is swept from $F_0$ to $F_0+\Delta F$ and then back to $F_0$:
\begin{equation}
F(t)=F_0+ \frac{t}{t_s}\Delta F\,\theta(t_s-t)
-\frac{t-2t_s}{t_s}\Delta F\,\theta(t-t_s),
\label{eq:4}
\end{equation}
where $\theta(t)$ denotes the Heaviside step function and $t_s$ is the sweep duration. For a coherent optical drive, the driving amplitude is related to the input power $P$ through $F=\sqrt{2\kappa_a P/\hbar\omega_l}$. 
The parameters $F_0$ and $\Delta F$ are chosen such that the sweep spans the full hysteresis region. Eqs. \eqref{lang_eqs1}-\eqref{lang_eqs2} define the open-system dynamics of the hybrid molecular optomechanical system and provide the starting point for the deterministic mean-field description introduced in the next section.


\section{Mean Field approximation}\label{sec3}

We define the quadrature operators for the plasmonic, cavity, and molecular-vibrational modes as
\begin{equation}
x_{\mathcal{O}}=\frac{\mathcal{O}^{\dagger}+\mathcal{O}}{\sqrt{2}},
\qquad
p_{\mathcal{O}}=i\frac{\mathcal{O}^{\dagger}-\mathcal{O}}{\sqrt{2}},
\end{equation}
where $\mathcal{O}\in\{a,c,B\}$, with $a$, $c$, and $B$ denoting the plasmonic cavity, FP cavity, and collective molecular-vibrational modes, respectively. Their expectation values are denoted by the corresponding capital letters, $X_{\mathcal{O}}=\langle x_{\mathcal{O}}\rangle$ and $P_{\mathcal{O}}=\langle p_{\mathcal{O}}\rangle$. Taking expectation values of Eqs. (\ref{lang_eqs1})-(\ref{lang_eqs2}) and using the zero-mean property of the input-noise operators, we obtain the equations governing the average amplitudes. Within the mean-field approximation, each operator is decomposed as $\mathcal{O}=\langle\mathcal{O}\rangle+\delta\mathcal{O}$, and correlations involving fluctuation operators are neglected, leading to the factorization $\langle AB\rangle\simeq\langle A\rangle\langle B\rangle$ \cite{PhysRevA.81.041803,Akram_2015,mean_field_mol_opt}. The resulting deterministic mean-field dynamics is therefore governed by the following set of nonlinear differential equations:

\begin{align}\label{le1}
\dot{X}_a &=-\kappa_a X_a + \left(\Delta_a 
- \sqrt{2}\,g_N X_B\right) P_a + J P_c + \sqrt{2}\,F(t),\nonumber\\
\dot{P}_a &=-\kappa_a P_a- \left(\Delta_a- \sqrt{2}\,g_N X_B\right) X_a- J X_c,\nonumber\\
\dot{X}_c &=-\kappa_c X_c + \Delta_c P_c + J P_a,\nonumber\\
\dot{P}_c &=-\kappa_c P_c- \Delta_c X_c- J X_a,\\
\dot{X}_B &=-\gamma_m X_B + \omega_m P_B,\nonumber\\
\dot{P}_B &=-\gamma_m P_B- \omega_m X_B
+ \frac{g_N}{\sqrt{2}}\left(X_a^2 + P_a^2\right).\nonumber\label{le2}
\end{align}
We investigate the dynamical hysteresis exhibited by this system through numerically integrating the deterministic mean-field equations over the interval $0\leq t\leq2t_s$ using Mathematica. The mean photon and phonon populations are evaluated as $\langle\mathcal O^\dagger\mathcal O\rangle\simeq|\langle\mathcal O\rangle|^2=(X_{\mathcal O}^2+P_{\mathcal O}^2)/2$. For the hysteresis protocol, the system is initially prepared in the zero-drive $F(0)=F_0=0$, corresponding to the unexcited configuration $X_{\mathcal O}(0)=P_{\mathcal O}(0)=0$. The driving amplitude is then linearly increased from zero to $\Delta F$ and subsequently decreased back to zero, following the triangular protocol defined in Eq. \eqref{eq:4}. The resulting nonlinear dynamics is obtained numerically, while the associated stationary solutions are analyzed in Sec. \ref{sec4}.

To demonstrate the experimental feasibility and practical applicability of the proposed memory effects, we consider experimentally accessible parameters in our numerical simulations. For the hybrid molecular COM system \cite{Shlesinger2023, Esteban2022, Roelli2016}, we use $\omega_a/2\pi=\omega_c/2\pi=460\,\text{THz}$, $P=1.15\,\text{mW}$, $\kappa_a/2\pi=4.6\,\text{THz}$, $\kappa_c/2\pi=46\,\text{GHz}$, $\omega_m/2\pi=20.7\,\text{THz}$, $\gamma_m/2\pi=4.6\,\text{GHz}$, $J/2\pi=44.6\,\text{THz}$, and $g/2\pi=70\,\text{GHz}$. Within the mean-field approximation, thermal fluctuations are neglected, and only deterministic dynamics is considered, as discussed in Sec. \ref{sec6}.

\begin{figure*}[t]
\centering
\includegraphics[width=0.95\linewidth]{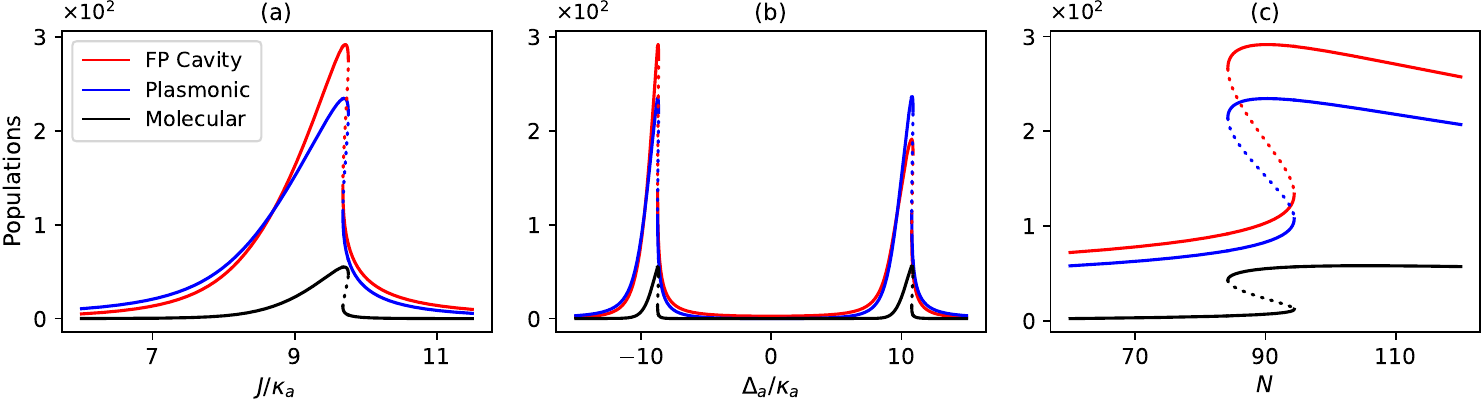}
\caption{Steady-state populations of the plasmonic cavity, FP cavity, and molecular vibrational modes at $F/\kappa_a=15.5$ as functions of (a) the coupling strength $J/\kappa_a$, (b) the plasmonic detuning $\Delta_a/\kappa_a$, and (c) the number of molecules $N$. Solid and dotted lines indicate stable and unstable steady-state branches, respectively. Other parameters are the same as in Fig. \ref{fig2}.}
\label{fig3}
\end{figure*}

\section{Molecular Optomechanics-Induced Cavity Memory}\label{sec4}

In this section, we characterize the hysteretic memory response of the hybrid molecular optomechanical system and analyze how it can be controlled through the relevant system parameters. We first examine the
dynamical hysteresis induced by a time-dependent driving field and then relate this behavior to the underlying bistable steady-state response.

\subsection{Dynamical Hysteresis}

In Fig. \ref{fig2}, we show the time-dependent populations of the plasmonic cavity $\langle \hat{a}^{\dagger}\hat{a}\rangle$, FP cavity $\langle \hat{c}^{\dagger}\hat{c}\rangle$, and molecular-vibrational $\langle \hat{B}^{\dagger}\hat{B}\rangle$ modes as functions of the driving amplitude $F(t)$. For clarity, only the driving-amplitude interval containing the hysteretic switching region is displayed, although the numerical evolution is performed over the full triangular driving protocol. All three modes exhibit a pronounced hysteretic response, characterized by distinct forward and backward branches and abrupt switching between low- and high-population states, as commonly observed in nonlinear optomechanical systems operating in the bistable regime \cite{Ghobadi2011,Aldana2013}. During the increasing sweep, the system follows the lower branch until a switching threshold is reached, where it jumps to the upper branch, remaining there during the decreasing sweep until the lower switching threshold is reached. Thus, for the same instantaneous value of $F(t)$, the system can occupy different dynamical states depending on its previous evolution, giving rise to a history-dependent, memory-like response. 

\subsection{Bistability of the steady-state solutions}\label{sec4b}

To identify the parameter regions supporting bistability and hysteresis, we examine the stationary solutions of the mean-field equations, Eqs. \eqref{le1}, and their linear stability. This analysis allows us to distinguish between stable and unstable branches and identify the coexistence regions underlying the hysteretic response. To this end, we consider a constant driving amplitude, $F(t)\equiv F$, and impose the steady-state conditions $\dot X_{\mathcal O}=\dot P_{\mathcal O}=0$. The resulting steady-state complex amplitudes satisfy:
\begin{align}
a_s&=\frac{X_{a,s}+iP_{a,s}}{\sqrt{2}},\\
c_s&=\frac{X_{c,s}+iP_{c,s}}{\sqrt{2}},\\
B_s&=\frac{X_{B,s}+iP_{B,s}}{\sqrt{2}}.
\end{align}
The stationary cavity and molecular equations can then be solved in terms of the plasmonic amplitude as
\begin{equation}
c_s=-\frac{iJ}{\kappa_c+i\Delta_c}a_s,
\qquad
B_s=\frac{i g_N |a_s|^2}{\gamma_m+i\omega_m},
\end{equation}
so that the cavity and molecular degrees of freedom can be eliminated from the stationary plasmonic equation (see Appendix \ref{app:bistability} for details). This leads to the nonlinear relation
\begin{equation}
F^2=n_a\left[\kappa_{\rm eff}^2+\left(\Delta_{\rm eff}-Kn_a\right)^2\right],
\label{eq:ssrelation}
\end{equation}
where $n_a=|a_s|^2$ and
\begin{equation}\label{effective_parameters}
\kappa_{\rm eff}=\kappa_a+\frac{J^2\kappa_c}{\kappa_c^2+\Delta_c^2},\qquad \Delta_{\rm eff}=\Delta_a-\frac{J^2\Delta_c}{\kappa_c^2+\Delta_c^2},
\end{equation}
while
\begin{equation}\label{Keff}
K=\frac{2g^2N\omega_m}{\omega_m^2+\gamma_m^2}.
\end{equation}
The corresponding cavity and collective molecular-vibrational populations are
\begin{equation}
n_c=\frac{J^2}{\kappa_c^2+\Delta_c^2}n_a,
\qquad
n_B=\frac{g^2N}{\omega_m^2+\gamma_m^2}n_a^2.
\end{equation}
For the present convention, where $K>0$, two physically relevant turning points exist when $\Delta_{\rm eff}>\sqrt{3}\kappa_{\rm eff}$ as derived in Appendix \ref{app:bistability}. The stability of the stationary solutions is determined using the Routh-Hurwitz criterion \cite{PhysRevA.35.5288}, with numerical evaluation for the parameters considered. Solid and dotted curves in Fig. \ref{fig3} indicate stable and unstable branches, respectively.

\begin{figure*}[t]
\centering
\includegraphics[width=0.8\linewidth]{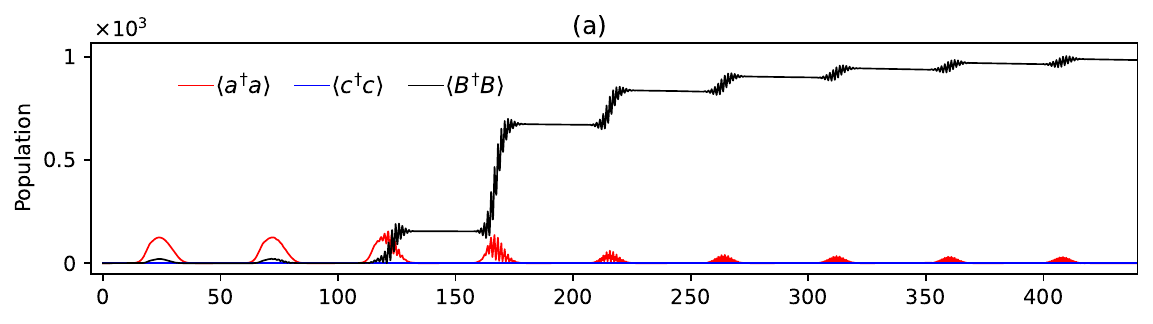}
\includegraphics[width=0.8\linewidth]{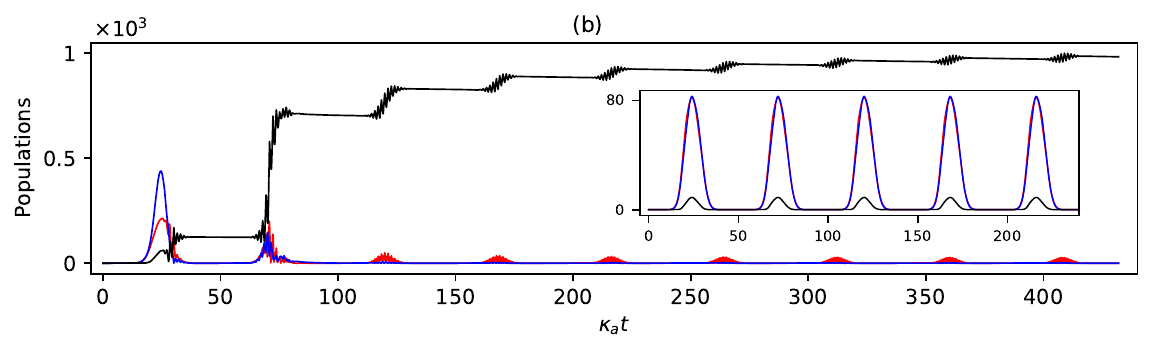}
\caption{Multilevel molecular switching under pulsed excitation. Temporal evolution of the plasmonic cavity, FP cavity, and collective molecular-vibrational populations for (a) $J=0$ and (b) $J=-\Delta_c=\kappa_a$. The collective molecular mode exhibits a stepwise response, with successive pulses driving the system toward distinct vibrational excitation levels. Coupling to the FP cavity enables higher molecular levels to be accessed at earlier times. The inset shows the suppressed response obtained for $J=\Delta_c=\kappa_a$. Other parameters (in units of $\kappa_a$): $\omega_m=4.5$, $\Delta_a=0$, $g=0.015$, $N=120$, $\kappa_c=10^{-2}$, $\gamma_m=10^{-4}$, $F_p=20$, $t_s=24/\kappa_a$, $\sigma=t_s/5$.}
\label{figm}
\end{figure*}

Figure \ref{fig3}\hyperref[fig3]{(a)} shows how the FP cavity-plasmon coupling, $J$ controls bistability. The upper and lower stable branches coexist over a finite range of coupling strengths, forming a bistable region characteristic of nonlinear optomechanical systems \cite{Ghobadi2011,Aldana2013}. According to Eq. \eqref{effective_parameters}, varying $J$ simultaneously renormalizes the effective optical linewidth and detuning, thereby providing a direct mechanism for controlling the bistable response. The FP cavity mode exhibits the largest population, followed by the plasmonic cavity and molecular-vibrational modes.

Figure \ref{fig3}\hyperref[fig3]{(b)} shows the corresponding populations as a function of the detuning $\Delta_a/\kappa_a$, with $\Delta_a=\Delta_c$. Pronounced nonlinear responses occur for both positive and negative bare detunings. Importantly, the relevant quantity governing the onset of bistability is the hybridized detuning $\Delta_{\mathrm{eff}}$, rather than the bare detuning $\Delta_a$ alone. For $\Delta_a<0$, the cavity population is strongly enhanced, whereas for $\Delta_a>0$, the plasmonic population becomes dominant. The molecular-vibrational mode follows the nonlinear response of the optical modes, demonstrating that the bistable response can be controlled through the optical detuning.

Finally, Fig. \ref{fig3}\hyperref[fig3]{(c)} shows the steady-state populations as a function of $N$. Since $g_N=g\sqrt{N}$, Eq. \eqref{Keff} gives $K\propto N$, showing that increasing the molecular ensemble size directly enhances the effective optomechanical nonlinearity. Consequently, both the position and width of the bistable region can be controlled through the collective molecular interaction.

\section{Controlled multilevel Molecular Switching}\label{sec5}

\subsection{Sequential multilevel molecular response}\label{sec5a}

Beyond bistability, we investigate the multilevel dynamics of the collective molecular vibrational mode under pulsed excitation. To this end, we consider a train of Gaussian pulses described by
\begin{equation}\label{pulsedriven}
    F(t)=\sum_{n\,\mathrm{odd}} F_p
    \exp\left[-\frac{(t-t_n)^2}{2\sigma^2}\right],
\end{equation}
where $F_p$ is the pulse amplitude, $t_n=n\,t_s$ is the pulse arrival time, and $\sigma$ is the pulse width. 

As shown in Fig. \ref{figm}, pulsed excitation produces a pronounced staircase-like response in the collective molecular-vibrational populations. For $\Delta_a=0$, successive pulses drive the collective vibrational mode toward progressively higher excitation plateaus. This behavior is particularly evident for $N=120$, $t_s=24/\kappa_a$, and $\sigma=t_s/5$. Each population step is accompanied by transient oscillations in the plasmonic populations, reflecting the rapid response of the coupled fields to the pulsed excitation. Such dynamics are reminiscent of amplitude-locking phenomena reported in Ref. \cite{PhysRevA.105.013521} and discrete phononic transitions predicted in pulsed optomechanical memory protocols \cite{Ariel2026}.

The role of the cavity-plasmon coupling in shaping the stepwise molecular response is shown in Fig. \ref{figm}\hyperref[figm]{(b)}. In contrast to the uncoupled case [Fig. \ref{figm}\hyperref[figm]{(a)}], for a blue-detuned cavity configuration with $\Delta_c=-\kappa_a$ and $J=\kappa_a$, the mechanical populations reach higher dynamical levels at earlier times. This enhancement is consistent with previous studies showing that a suitable combination of blue detuning and intercavity coupling can promote phonon amplification \cite{Xie:20}. In our system, the FP cavity therefore acts as an additional control channel for the driven vibrational response, reducing the timescale required to reach a given excitation level. Remarkably, this behavior is highly sensitive to the cavity detuning. As shown in the inset of Fig. \ref{figm}\hyperref[figm]{(b)}, reversing the detuning strongly suppresses the multilevel molecular response.

A detailed analysis of the first-pulse molecular population as a function of the cavity-plasmon coupling and molecular ensemble size is provided in Appendix \ref{app:firstpulse}.

\begin{figure*}[t]
\centering
\includegraphics[width=0.8\linewidth]{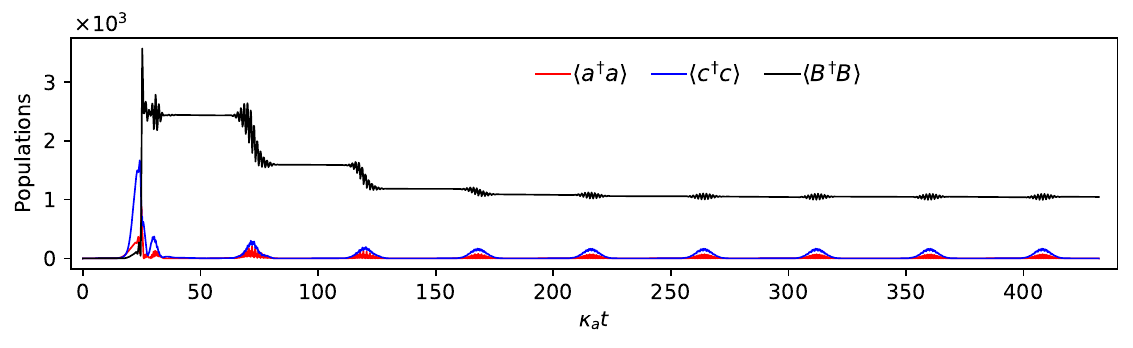}
\caption{Descending vibrational staircase under simultaneous pulsed excitation of the plasmonic and FP cavity modes. Both optical modes are driven by synchronized Gaussian pulse trains with $F_c/\kappa_a=12$. The first pulse generates a large vibrational excitation, whereas subsequent pulses produce progressively smaller downward transitions between population plateaus. Other parameters are the same as in Fig.  \ref{figm}\hyperref[figm]{(b)}.}
\label{fig2p}
\end{figure*}

\subsection{Pulse-induced vibrational depletion}
\label{sec:dual_drive_depletion}

We extend the pulsed protocol by driving the auxiliary FP cavity in addition to the plasmonic mode. This introduces an additional excitation channel into the coupled optical system, allowing us to investigate how simultaneous driving of both optical modes influences the vibrational response. Assuming that both driving fields have the same frequency $\omega_l$ and zero relative phase, the additional Hamiltonian driving term in the rotating frame at $\omega_l$ reads
\begin{equation}
 H_{c}=iF_c(t)(c^\dagger-c),
 \label{eq:dd_hdrive}
\end{equation}
where $F_c(t)$ follows the pulsed driving protocol defined in Eq. \eqref{pulsedriven}, with pulse amplitude $F_c$. 

Following the numerical procedure described in Sec. \ref{sec5a}, we investigate the dynamics governed by the total Hamiltonian $H+H_c$. Figure \ref{fig2p} shows that simultaneous driving produces a descending vibrational staircase, where the first pulse generates a large molecular excitation and subsequent pulses induce stepwise population depletion. Each Gaussian pulse produces a transient plasmonic excitation that modifies the radiation-pressure force acting on the molecular vibrations. Since $\gamma_m\ll\kappa_a$, vibrational excitation persists between pulses, resulting in nearly constant population plateaus.

The staircase direction depends on the driving configuration and the vibrational state prepared by previous pulses. With plasmonic driving alone, successive pulses increase the molecular population [Fig. \ref{figm}]. Adding the auxiliary-cavity drive $F_c$ enhances the initial excitation and modifies the plasmonic response. For $J=-\Delta_c$, intercavity photon exchange redistributes the optical populations, allowing subsequent pulses to partially suppress the accumulated vibrational excitation. This behavior is reminiscent of interference-based optomechanical control \cite{Kuzyk2017} and staircase-like mechanical responses under multiple optical drives \cite{Xiao2026}. Thus, the driving amplitudes, intercavity coupling, and pulse timing provide control over the direction and magnitude of vibrational population transitions.

\section{Tunability and Experimental Feasibility}\label{sec6}

The parameter regime considered in our simulations is compatible with characteristic frequency and coupling scales reported for molecular optomechanical platforms. Molecular vibrations can span a broad range of frequencies, approximately $\omega_m/2\pi=0.749$-$48.3\,\text{THz}$ \cite{Roelli2016}, while strong single-molecule optomechanical coupling strengths can be achieved in plasmonic nanocavities due to their extremely small effective mode volumes. For representative NPoM configurations, coupling strengths in the range $g/2\pi\simeq0.0295$-$0.628\,\text{THz}$ have been reported for effective mode volumes of the order of $V_m\simeq1500\,\mathrm{nm}^{3}$ \cite{Roelli2016}. These values include molecular and solid-state vibrational resonances associated with systems such as R6G, GBT, graphene, and the G band of carbon nanotubes \cite{10.1021/jp045771u,PhysRevB.82.045418}, providing possible candidates for accessing the parameter regimes explored here.

For comparison with experimentally relevant values, the dimensionless parameters employed in our calculations can be related to a plasmonic linewidth $\kappa_a/2\pi=4.6\,\text{THz}$. For instance, $\omega_m/\kappa_a=4.5$ and $g/\kappa_a=0.015$ correspond to $\omega_m/2\pi\simeq20.7\,\text{THz}$ and $g/2\pi\simeq70\,\text{GHz}$, respectively, while $\kappa_c/\kappa_a=10^{-2}$ corresponds to a cavity linewidth $\kappa_c/2\pi\simeq0.046\,\text{THz}$. The cavity-plasmon coupling used in the main analysis, $J/\kappa_a=9.7$, corresponds to $J/2\pi\simeq44.6\,\text{THz}$. Strong plasmon-cavity hybridization on this scale is experimentally accessible. In particular, coupling energies of approximately $300\,\text{meV}$, corresponding to coupling frequencies of about $72\,\text{THz}$, have been experimentally extracted for plasmonic nanorod arrays embedded in FP microcavities \cite{Baranov2020}. These results support the optical coupling scale considered here, while molecular optomechanical platforms provide vibrational frequencies and single-molecule coupling strengths within the ranges employed in our calculations \cite{mean_field_mol_opt,Shlesinger2023,Roelli2016}. 

\begin{figure}[t]
\centering
\includegraphics[width=1\linewidth]{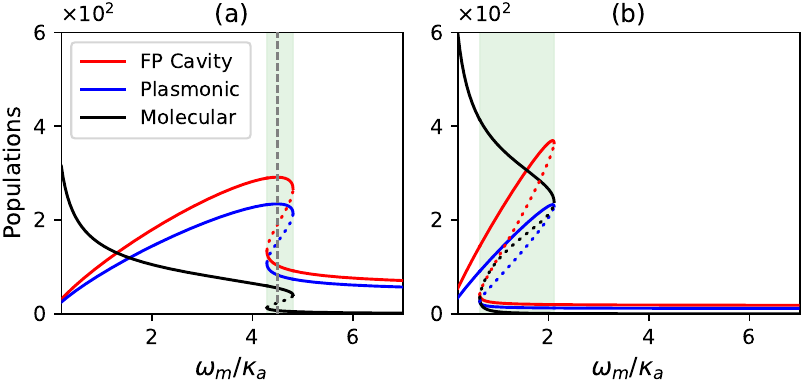}
\caption{Control of the bistable response at $F/\kappa_a=15.5$ as a function of the mechanical frequency: $(a)$ $\Delta_a=-8.7$ and $(b)$ $\Delta_a=-7.7$. For the hybrid COM system, increasing $\Delta_a$ shifts the bistable response toward lower mechanical frequencies. The shaded region indicates the parameter interval in which multiple stationary solutions coexist, while the dashed line marks the mechanical frequency considered in the main analysis. Other parameters are the same as in Fig. \ref{fig2}.}
\label{figw}
\end{figure}

Importantly, the emergence of bistability and hysteresis is not restricted to a single molecular vibrational frequency. Figure \ref{figw} illustrates the bistable response as a function of $\omega_m/\kappa_a$ for two different optical detunings. For $\Delta_a/\kappa_a=-8.7$, the bistable regime occurs around the vibrational frequency considered in the main analysis. Increasing $\Delta_a/\kappa_a$ to $-7.7$ shifts the bistable region toward lower mechanical frequencies. This behavior shows that variations in the molecular vibrational frequency can be partially compensated by tuning the optical detuning, thereby extending the range of molecular resonances for which the hysteretic response can be accessed. Such tunability is particularly relevant for molecular optomechanical implementations, where different molecules naturally possess different characteristic vibrational frequencies.

For the representative optical parameters considered here, with $\omega_a/2\pi\simeq\omega_c/2\pi\simeq460\,\text{THz}$, the driving amplitudes employed in both the hysteresis and pulsed vibrational staircase protocols correspond to optical powers on the $\text{mW}$ scale. This places the predicted nonlinear dynamics within a parameter regime relevant to plasmonic molecular optomechanical platforms.

Regarding thermal effects, the reservoir fluctuations are formally included through the input-noise operators in the Heisenberg-Langevin equations. However, the present mean-field treatment focuses on the deterministic dynamics of the average amplitudes. Since the input-noise operators have vanishing first moments, they do not contribute explicitly to the mean-field equations, although their correlation functions depend on the reservoir temperature and become relevant for the fluctuation dynamics. At the optical frequencies of the plasmonic and cavity modes, $\omega_{a,c}/2\pi\simeq460\,\mathrm{THz}$, the corresponding thermal photon occupations are negligible at room temperature, and the optical reservoirs can therefore be effectively treated as vacuum baths \cite{Aspelmeyer2014,PhysRevA.77.033804}. Temperature is more relevant for the molecular vibrational mode through its thermal phonon occupation, $\bar{n}_{\mathrm{th}}=[\exp(\hbar\omega_m/k_BT)-1]^{-1}$. For the vibrational frequency considered here, $\omega_m/2\pi=20.7\,\mathrm{THz}$, this occupation remains small even at room temperature \cite{Esteban2022,Roelli2016,Roelli_Review2024}. Furthermore, possible photothermal effects arising from optical absorption in the metallic nanocavity \cite{Thermo-plasmonics2013} are beyond the scope of the present model.

\vspace{-0.6cm}
\section{Conclusions}\label{sec7}
\vspace{-0.2cm}

In this work, we have investigated the nonlinear dynamics of a hybrid molecular optomechanical system comprising a plasmonic nanocavity, a FP cavity, and an ensemble of molecular vibrations. We have demonstrated that molecular optomechanical interactions and cavity-plasmon coupling give rise to bistability and dynamical hysteresis in the optical and molecular-vibrational populations. The auxiliary FP cavity modifies the effective optical linewidth and detuning, while collective molecular interactions determine the strength of the nonlinearity. Consequently, the bistable regime and the associated history-dependent response can be controlled through the cavity-plasmon coupling, optical detuning, and molecular ensemble size.

Beyond bistability, we have investigated the pulse-driven dynamics of the collective molecular vibrational mode. Pulsed excitation produces ascending or descending vibrational staircases depending on the optical driving configuration. The auxiliary FP cavity controls the buildup of vibrational excitation through intercavity coupling and detuning, while simultaneous driving of both optical modes enables stepwise vibrational depletion. These results establish hybrid molecular optomechanical systems as a tunable platform for controlling history-dependent and pulse-driven molecular dynamics, with potential applications in nanoscale memory, multilevel information processing, and optically controlled vibrational switching.

\section*{Acknowledgments}

H.M. acknowledges the Universidad San Sebastián through the postdoctoral fellowship.
E.C. acknowledges the support of the Fondo Nacional
de Desarrollo Científico y Tecnológico (FONDECYT, ANID-Chile) through Project No. 1231018. Institutional support provided by Universidad San Sebastián through the initiative Grant No. USS-FIN-25-FIAC-02 is also gratefully acknowledged.

\appendix

\section{Analytical derivation of the steady-state bistability}\label{app:bistability}

In this Appendix, we derive the nonlinear steady-state relation used in Sec. \ref{sec4} from the mean-field equations of motion, Eqs. \eqref{le1}. We introduce the complex mean-field amplitudes
\begin{equation}
a=\frac{X_a+iP_a}{\sqrt{2}},\quad
c=\frac{X_c+iP_c}{\sqrt{2}},\quad
B=\frac{X_B+iP_B}{\sqrt{2}},
\end{equation}
so that the corresponding stationary amplitudes are denoted by $a_s$, $c_s$, and $B_s$. Combining Eqs. \eqref{le1} into complex form and setting all time derivatives to zero gives
\begin{align}
0&=-(\kappa_a+i\Delta_a)a_s-iJc_s
+i g_N a_s(B_s+B_s^*)+F,\\
0&=-(\kappa_c+i\Delta_c)c_s-iJa_s,\\
0&=-(\gamma_m+i\omega_m)B_s+i g_N|a_s|^2.
\end{align}
The last two equations can be solved directly, yielding
\begin{equation}
c_s=-\frac{iJ}{\kappa_c+i\Delta_c}a_s,
\qquad
B_s=\frac{i g_N|a_s|^2}{\gamma_m+i\omega_m}.
\end{equation}
Consequently,
\begin{equation}
B_s+B_s^{*}=\frac{2g_N\omega_m}{\omega_m^2+\gamma_m^2}|a_s|^2.
\label{app:Bshift}
\end{equation}
Substituting these expressions into the stationary plasmonic equation yields
\begin{equation}
a_s=\frac{F}{\kappa_{\mathrm{eff}}+i\left(\Delta_{\mathrm{eff}}-Kn_a\right)},
\label{app:as}
\end{equation}
where $n_a=|a_s|^2$ and
\begin{equation}
\kappa_{\mathrm{eff}}=\kappa_a+\frac{J^2\kappa_c}{\kappa_c^2+\Delta_c^2},
\qquad
\Delta_{\mathrm{eff}}=\Delta_a-\frac{J^2\Delta_c}
{\kappa_c^2+\Delta_c^2},
\end{equation}
while
\begin{align}
K&=\frac{2g_N^2\omega_m}{\omega_m^2+\gamma_m^2}.
\end{align}
Taking the modulus squared of Eq. \eqref{app:as} gives
\begin{equation}\label{f2}
F^2=n_a\left[\kappa_{\mathrm{eff}}^2+
\left(\Delta_{\mathrm{eff}}-Kn_a\right)^2
\right],
\end{equation}
or equivalently
\begin{equation}
K^2n_a^3-2K\Delta_{\mathrm{eff}}n_a^2
+\left(\kappa_{\mathrm{eff}}^2+\Delta_{\mathrm{eff}}^2\right)n_a-F^2=0.
\label{app:cubic}
\end{equation}
Equation \eqref{app:cubic} reveals an intensity-dependent optical detuning arising from the collective molecular optomechanical interaction. The FP cavity, in turn, renormalizes both the effective linewidth and detuning. The remaining stationary populations are obtained directly as
\begin{equation}
n_c=\frac{J^2}{\kappa_c^2+\Delta_c^2}\,n_a,\quad
n_B=\frac{g_N^2}{\omega_m^2+\gamma_m^2}\,n_a^2.
\end{equation}
The nonlinear relation, Eq. \eqref{f2}, defines the driving strength required to sustain a stationary plasmonic population $n_a$. A multivalued response emerges when $F^2(n_a)$ develops two turning points, corresponding to the folds of the stationary curve. These points are obtained from $dF^2/dn_a=0$, which gives
\begin{equation}
3K^2n_a^2-4K\Delta_{\rm eff}n_a+\left(\Delta_{\rm eff}^2+\kappa_{\rm eff}^2\right)=0.
\end{equation}
Solving this quadratic equation yields the populations at the two turning points
\begin{equation}
n_{\pm}=\frac{2\Delta_{\rm eff}\pm\sqrt{\Delta_{\rm eff}^2-3\kappa_{\rm eff}^2}}{3K}.
\end{equation}
In addition, the corresponding populations must satisfy $n_{\pm}>0$. Since $K>0$ in the present system, this condition selects
\begin{equation}
\Delta_{\rm eff}>\sqrt{3}\kappa_{\rm eff}.
\end{equation}
Under this condition, the stationary curve can develop an S-shaped structure, leading to a range of driving amplitudes over which three stationary plasmonic populations coexist, as discussed in Sec. \ref{sec4b}.

\vspace{0.6cm}
\section{Control of the first-pulse molecular response}\label{app:firstpulse}

\begin{figure}[ht]
\centering
\includegraphics[width=1\linewidth]{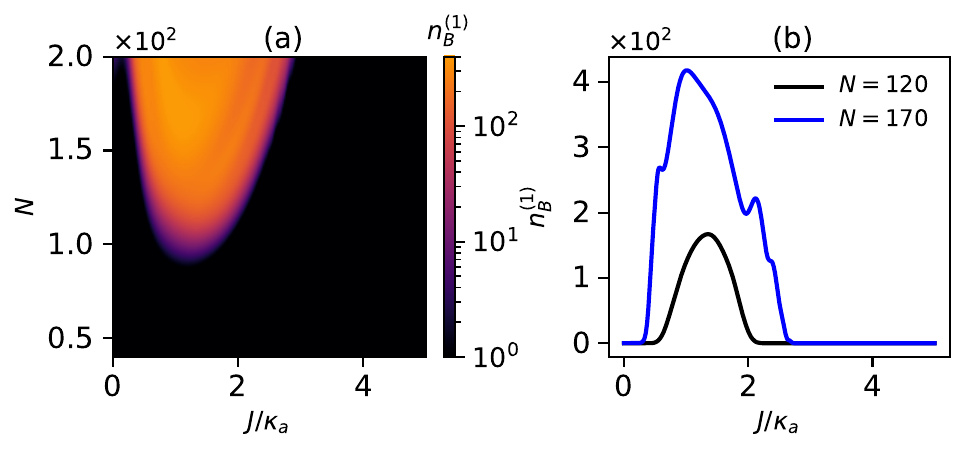}
\caption{First-pulse molecular population as a function of the cavity-plasmon coupling and molecular ensemble size. (a) $n_B^{(1)}$ as a function of $J/\kappa_a$ and $N$, for $\Delta_c=-J$. (b) Cuts of panel (a) for $N=120$ and $170$. Other parameters are the same as in Fig. \ref{figm}.}
\label{figmc}
\end{figure}

To further characterize the control of the first accessible molecular level, we analyze the population reached after the first pulse as a function of the cavity-plasmon coupling $J$ and the number of
molecules $N$. We define the total collective molecular-vibrational population after the first pulse as $n_B^{(1)}\equiv\langle B^\dagger B\rangle
|_{t=2t_s}$, where $t=2t_s$ corresponds to the midpoint between the first and second pulses. Figure \ref{figmc}\hyperref[figmc]{(a)} shows $n_B^{(1)}$ under the condition $\Delta_c=-J$. A pronounced molecular response emerges for sufficiently large $N$, reflecting the collective enhancement of the molecular optomechanical interaction.

The cuts shown in Fig. \ref{figmc}\hyperref[figmc]{(b)} reveal an optimal coupling range in which the first-pulse molecular population is maximized. Along the condition $\Delta_c=-J$, the vibrational excitation initially increases with $J$, reaches a maximum at intermediate coupling strengths, and subsequently decreases as the coupling becomes stronger. This nonmonotonic behavior demonstrates that the first-pulse molecular response can be optimized through the cavity-plasmon coupling and detuning.

\bibliography{biblio}

@article{Baranov2020,
  author = {Baranov, Denis G. and Munkhbat, Battulga and Zhukova, Elena and Bisht, Ankit and Canales, Adriana and Rousseaux, Benjamin and Johansson, G{\"o}ran and Antosiewicz, Tomasz J. and Shegai, Timur},
  title = {Ultrastrong coupling between nanoparticle plasmons and cavity photons at ambient conditions},
  journal = {Nat. Commun.},
  volume = {11},
  pages = {2715},
  year = {2020},
  doi = {10.1038/s41467-020-16524-x}
}

@article{PhysRevA.105.013521,
  title = {Amplitude and phase locking of mechanical oscillation driven by radiation pressure},
  author = {Wu, Yi and Li, Gang and He, Bing and Lin, Qing},
  journal = {Phys. Rev. A},
  volume = {105},
  issue = {1},
  pages = {013521},
  numpages = {12},
  year = {2022},
  month = {Jan},
  publisher = {American Physical Society},
  doi = {10.1103/PhysRevA.105.013521},
  url = {https://link.aps.org/doi/10.1103/PhysRevA.105.013521}
}

@article{Ashrafi2019,
  title = {Optomechanical heat transfer between molecules in a nanoplasmonic cavity},
  author = {Ashrafi, S. Mahmoud and Malekfar, R. and Bahrampour, A. R. and Feist, Johannes},
  journal = {Phys. Rev. A},
  volume = {100},
  issue = {1},
  pages = {013826},
  numpages = {9},
  year = {2019},
  month = {7},
  publisher = {American Physical Society},
  doi = {10.1103/PhysRevA.100.013826},
  url = {https://link.aps.org/doi/10.1103/PhysRevA.100.013826}
}

@article{Deeg2025,
  author  = {L. F. Deeg and D. Zoepfl and N. Diaz-Naufal and M. L. Juan and A. Metelmann and G. Kirchmair},
  title   = {Optomechanical backaction in the bistable regime},
  journal = {Phys. Rev. Applied},
  volume  = {23},
  pages   = {014082},
  year    = {2025},
  doi     = {10.1103/PhysRevApplied.23.014082}
}

@article{She2025,
  author  = {Yanchao She and Mingqi Xu and Shixia Xiang and Chenyu Wu and Hongyu Xiao and Weixi Zhang},
  title   = {Optical bistable switching in a dual-cavity hybrid optomechanical system filled with {K}err media and semiconductor double quantum dot molecules},
  journal = {Phys. Lett. A},
  volume  = {561},
  pages   = {130976},
  year    = {2025},
  doi     = {10.1016/j.physleta.2025.130976}
}

@article{Ariel2026,
  title = {Memory effects in pulsed optomechanical systems},
  author = {Tapia-Maureira, Hachisko and He, Bing and Di Ventra, Massimiliano and Norambuena, Ariel},
  journal = {Phys. Rev. Appl.},
  volume = {25},
  issue = {4},
  pages = {044083},
  numpages = {13},
  year = {2026},
  month = {Apr},
  publisher = {American Physical Society},
  doi = {10.1103/mxbn-15ry},
  url = {https://link.aps.org/doi/10.1103/mxbn-15ry}
}

@article{npom_adm,
author = {Li, Yang and Chen, Wen and He, Xiaobo and Shi, Junjun and Cui, Ximin and Sun, Jiawei and Xu, Hongxing},
title = {Boosting Light-Matter Interactions in Plasmonic Nanogaps},
journal = {Advanced Materials},
volume = {36},
number = {49},
pages = {2405186},
doi = {https://doi.org/10.1002/adma.202405186},
url = {https://advanced.onlinelibrary.wiley.com/doi/abs/10.1002/adma.202405186},
year = {2024}
}

@article{npom2022,
url = {https://doi.org/10.1515/nanoph-2022-0521},
title = {Nanoparticle-on-mirror pairs: building blocks for remote spectroscopies},
author = {Huatian Hu and Yuhao Xu and Zhiwei Hu and Bowen Kang and Zhenglong Zhang and Jiawei Sun and Yang Li and Hongxing Xu},
pages = {5153--5163},
volume = {11},
number = {22},
journal = {Nanophotonics},
doi = {doi:10.1515/nanoph-2022-0521},
year = {2022},
lastchecked = {2026-08-02}
}

@article{Aldana2013,
  title = {Equivalence between an optomechanical system and a {K}err medium},
  author = {Aldana, Samuel and Bruder, Christoph and Nunnenkamp, Andreas},
  journal = {Phys. Rev. A},
  volume = {88},
  issue = {4},
  pages = {043826},
  numpages = {10},
  year = {2013},
  month = {Oct},
  publisher = {American Physical Society},
  doi = {10.1103/PhysRevA.88.043826},
  url = {https://link.aps.org/doi/10.1103/PhysRevA.88.043826}
}

@article{Ghobadi2011,
  title = {Quantum optomechanics in the bistable regime},
  author = {Ghobadi, R. and Bahrampour, A. R. and Simon, C.},
  journal = {Phys. Rev. A},
  volume = {84},
  issue = {3},
  pages = {033846},
  numpages = {7},
  year = {2011},
  month = {Sep},
  publisher = {American Physical Society},
  doi = {10.1103/PhysRevA.84.033846},
  url = {https://link.aps.org/doi/10.1103/PhysRevA.84.033846}
}

@article{PhysRevA.77.033804,
  title = {Ground-state cooling of a micromechanical oscillator: Comparing cold damping and cavity-assisted cooling schemes},
  author = {Genes, C. and Vitali, D. and Tombesi, P. and Gigan, S. and Aspelmeyer, M.},
  journal = {Phys. Rev. A},
  volume = {77},
  issue = {3},
  pages = {033804},
  numpages = {9},
  year = {2008},
  month = {3},
  publisher = {American Physical Society},
  doi = {10.1103/PhysRevA.77.033804},
  url = {https://link.aps.org/doi/10.1103/PhysRevA.77.033804}
}

@article{Roelli_Review2024,
author = {Roelli, Philippe and Hu, Huatian and Verhagen, Ewold and Reich, Stephanie and Galland, Christophe},
title = {Nanocavities for Molecular Optomechanics: Their Fundamental Description and Applications},
journal = {ACS Photonics},
volume = {11},
number = {11},
pages = {4486-4501},
year = {2024},
doi = {10.1021/acsphotonics.4c01548},

URL = {https://doi.org/10.1021/acsphotonics.4c01548}
}

@article{Mol_blockade,
  author  = {Jian Tang and Baijun Li and Bin Yin and Tian-Xiang Lu and Ran Huang and Franco Nori and Hui Jing},
  title   = {Robust photon blockade with hybrid molecular optomechanics},
  journal = {npj Quantum Information},
  volume  = {12},
  number  = {1},
  pages   = {78},
  year    = {2026},
  doi     = {10.1038/s41534-026-01220-3},
}

@article{Gardiner1985,
  title = {Input and output in damped quantum systems: Quantum stochastic differential equations and the master equation},
  author = {Gardiner, C. W. and Collett, M. J.},
  journal = {Phys. Rev. A},
  volume = {31},
  issue = {6},
  pages = {3761--3774},
  numpages = {0},
  year = {1985},
  month = {Jun},
  publisher = {American Physical Society},
  doi = {10.1103/PhysRevA.31.3761},
  url = {https://link.aps.org/doi/10.1103/PhysRevA.31.3761}
}

@article{Wang2012,
  author  = {Y.-D. Wang and A. A. Clerk},
  title   = {Using Interference for High Fidelity Quantum State Transfer in Optomechanics},
  journal = {Phys. Rev. Lett.},
  volume  = {108},
  pages   = {153603},
  year    = {2012},
  doi     = {10.1103/PhysRevLett.108.153603}
}

@article{Vitali2007,
  title = {Optomechanical Entanglement between a Movable Mirror and a Cavity Field},
  author = {Vitali, D. and Gigan, S. and Ferreira, A. and B{\"o}hm, H. R. and Tombesi, P. and Guerreiro, A. and Vedral, V. and Zeilinger, A. and Aspelmeyer, M.},
  journal = {Phys. Rev. Lett.},
  volume = {98},
  issue = {3},
  pages = {030405},
  numpages = {4},
  year = {2007},
  month = {Jan},
  publisher = {American Physical Society},
  doi = {10.1103/PhysRevLett.98.030405},
  url = {https://link.aps.org/doi/10.1103/PhysRevLett.98.030405}
}

@article{Mol_phonon_laser,
author = {Yin, Bin and Wang, Jie and Zhang, Qian and Wang, Deng and Lu, Tian-Xiang and Jing, Hui},
title = {Molecular-Optomechanical Phonon Laser},
journal = {Laser \& Photonics Reviews},
volume = {},
number = {},
pages = {e71191},
year = {2026},
doi = {https://doi.org/10.1002/lpor.71191},
url = {https://onlinelibrary.wiley.com/doi/abs/10.1002/lpor.71191}
}

@article{Aspelmeyer2014,
  title = {Cavity optomechanics},
  author = {Aspelmeyer, Markus and Kippenberg, Tobias J. and Marquardt, Florian},
  journal = {Rev. Mod. Phys.},
  volume = {86},
  issue = {4},
  pages = {1391--1452},
  numpages = {62},
  year = {2014},
  month = {12},
  publisher = {American Physical Society},
  doi = {10.1103/RevModPhys.86.1391},
  url = {https://link.aps.org/doi/10.1103/RevModPhys.86.1391}
}

@article{Jakob2023,
  author  = {Jakob, Lukas A. and Deacon, William M. 
             and Zhang, Yuan and de Nijs, Bart 
             and Pavlenko, Elena and Hu, Shu 
             and Carnegie, Cloudy and Neuman, Tomas 
             and Esteban, Ruben and Aizpurua, Javier 
             and Baumberg, Jeremy J.},
  title   = {Giant optomechanical spring effect in plasmonic
             nano- and picocavities probed by
             surface-enhanced {R}aman scattering},
  journal = {Nat. Commun.},
  volume  = {14},
  pages   = {3291},
  year    = {2023},
  doi     = {10.1038/s41467-023-38124-1}
}

@article{opt_coup,
author = {Felix Benz  and Mikolaj K. Schmidt  and Alexander Dreismann  and Rohit Chikkaraddy  and Yao Zhang  and Angela Demetriadou  and Cloudy Carnegie  and Hamid Ohadi  and Bart de Nijs  and Ruben Esteban  and Javier Aizpurua  and Jeremy J. Baumberg },
title = {Single-molecule optomechanics in ``picocavities''},
journal = {Science},
volume = {354},
number = {6313},
pages = {726-729},
year = {2016},
doi = {10.1126/science.aah5243},
URL = {https://www.science.org/doi/abs/10.1126/science.aah5243}}

@article{Xie:20,
author = {Yi Fei Xie and Zhen Cao and Bing He and Qing Lin},
journal = {Opt. Express},
number = {15},
pages = {22580--22593},
publisher = {Optica Publishing Group},
title = {{PT}-symmetric phonon laser under gain saturation effect},
volume = {28},
month = {Jul},
year = {2020},
url = {https://opg.optica.org/oe/abstract.cfm?URI=oe-28-15-22580},
doi = {10.1364/OE.396893},
}

@article{Thermo-plasmonics2013,
author = {Baffou, Guillaume and Quidant, Romain},
title = {Thermo-plasmonics: using metallic nanostructures as nano-sources of heat},
journal = {Laser \& Photonics Reviews},
volume = {7},
number = {2},
pages = {171-187},
doi = {https://doi.org/10.1002/lpor.201200003},
url = {https://onlinelibrary.wiley.com/doi/abs/10.1002/lpor.201200003},
year = {2013}
}

@Article{Roelli2016,
  author  = {Philippe Roelli and Christophe Galland and Nicolas Piro and Tobias J. Kippenberg},
  title   = {Molecular cavity optomechanics as a theory of plasmon-enhanced {R}aman scattering},
  journal = {Nat. Nanotechnol.},
  year    = {2016},
  volume  = {11},
  number  = {2},
  pages   = {164--169},
  doi     = {10.1038/nnano.2015.264}
}

@Article{Esteban2022,
  author  = {Ruben Esteban and Jeremy J. Baumberg and Javier Aizpurua},
  title   = {Molecular Optomechanics Approach to Surface-Enhanced {R}aman Scattering},
  journal = {Acc. Chem. Res.},
  year    = {2022},
  volume  = {55},
  number  = {14},
  pages   = {1889--1899},
  doi     = {10.1021/acs.accounts.1c00759}
}

@article{Yin2025,
  title = {Molecular optomechanically induced transparency},
  author = {Yin, Bin and Wang, Jie and Peng, Mei-Yu and Zhang, Qian and Wang, Deng and Lu, Tian-Xiang and Wei, Ke and Jing, Hui},
  journal = {Phys. Rev. A},
  volume = {111},
  issue = {4},
  pages = {043507},
  numpages = {12},
  year = {2025},
  month = {4},
  publisher = {American Physical Society},
  doi = {10.1103/PhysRevA.111.043507},
  url = {https://link.aps.org/doi/10.1103/PhysRevA.111.043507}
}

@article{Xomalis2021,
author = {Angelos Xomalis  and Xuezhi Zheng  and Rohit Chikkaraddy  and Zsuzsanna Koczor-Benda  and Ermanno Miele  and Edina Rosta  and Guy A. E. Vandenbosch  and Alejandro Mart{\'i}nez  and Jeremy J. Baumberg },
title = {Detecting mid-infrared light by molecular frequency upconversion in dual-wavelength nanoantennas},
journal = {Science},
volume = {374},
number = {6572},
pages = {1268-1271},
year = {2021},
doi = {10.1126/science.abk2593},
URL = {https://www.science.org/doi/abs/10.1126/science.abk2593}}

@article{Zou2024,
  title = {Amplifying Frequency Up-Converted Infrared Signals with a Molecular Optomechanical Cavity},
  author = {Zou, Fen and Du, Lei and Li, Yong and Dong, Hui},
  journal = {Phys. Rev. Lett.},
  volume = {132},
  issue = {15},
  pages = {153602},
  numpages = {6},
  year = {2024},
  month = {4},
  publisher = {American Physical Society},
  doi = {10.1103/PhysRevLett.132.153602},
  url = {https://link.aps.org/doi/10.1103/PhysRevLett.132.153602}
}

@article{Liu2018a,
author = {Jian Liu and Ka-Di Zhu},
journal = {Photon. Res.},
number = {9},
pages = {867--874},
publisher = {Optica Publishing Group},
title = {Room temperature optical mass sensor with an artificial molecular structure based on surface plasmon optomechanics},
volume = {6},
month = {9},
year = {2018},
url = {https://opg.optica.org/prj/abstract.cfm?URI=prj-6-9-867},
doi = {10.1364/PRJ.6.000867}
}

@article{Tabatabaei2015,
    author = {Tabatabaei, Mohammadali and Najiminaini, Mohamadreza and Davieau, Kieffer and Kaminska, Bozena and Singh, Mahi R. and Carson, Jeffrey J. L. and Lagugn{\'e}-Labarthet, Fran{\c{c}}ois},
    title = {Tunable {3D} Plasmonic Cavity Nanosensors for Surface-Enhanced {R}aman Spectroscopy
with Sub-femtomolar Limit of Detection},
    journal = {ACS Photonics},
    volume = {2},
    number = {6},
    pages = {752-759},
    year = {2015},
    month = {5},
    issn = {2330-4022},
    doi = {10.1021/acsphotonics.5b00104},
}

@article{PhysRevB.82.045418,
  title = {Absolute {R}aman matrix elements of graphene and graphite},
  author = {Narula, Rohit and Panknin, Robert and Reich, Stephanie},
  journal = {Phys. Rev. B},
  volume = {82},
  issue = {4},
  pages = {045418},
  numpages = {7},
  year = {2010},
  month = {Jul},
  publisher = {American Physical Society},
  doi = {10.1103/PhysRevB.82.045418},
  url = {https://link.aps.org/doi/10.1103/PhysRevB.82.045418}
}

@article{10.1021/jp045771u,
    author = {Watanabe, Hiroyuki and Hayazawa, Norihiko and Inouye, Yasushi and Kawata, Satoshi},
    title = {{DFT} Vibrational Calculations of Rhodamine 6G Adsorbed on Silver: Analysis of
Tip-Enhanced {R}aman Spectroscopy},
    journal = {The Journal of Physical Chemistry B},
    volume = {109},
    number = {11},
    pages = {5012-5020},
    year = {2005},
    month = {03},
    issn = {1520-6106},
    doi = {10.1021/jp045771u},
    url = {https://doi.org/10.1021/jp045771u},
}

@article{Akram_2015,
doi = {10.1088/0953-4075/48/6/065502},
url = {https://doi.org/10.1088/0953-4075/48/6/065502},
year = {2015},
month = {feb},
publisher = {IOP Publishing},
volume = {48},
number = {6},
pages = {065502},
author = {Akram, Muhammad Javed and Ghafoor, Fazal and Saif, Farhan},
title = {Electromagnetically induced transparency and tunable fano resonances in hybrid optomechanics},
journal = {Journal of Physics B: Atomic, Molecular and Optical Physics}
}

@article{mean_field_mol_opt,
    author = {Shlesinger, Ilan and Cogn{\'{e}}e, K{\'{e}}vin G. and Verhagen, Ewold and Koenderink, A. Femius},
    title = {Integrated Molecular Optomechanics with Hybrid Dielectric--Metallic
Resonators},
    journal = {ACS Photonics},
    volume = {8},
    number = {12},
    pages = {3506-3516},
    year = {2021},
    month = {11},
    issn = {2330-4022},
    doi = {10.1021/acsphotonics.1c00808},
    url = {https://doi.org/10.1021/acsphotonics.1c00808}
}

@article{PhysRevA.81.041803,
  title = {Electromagnetically induced transparency in mechanical effects of light},
  author = {Agarwal, G. S. and Huang, Sumei},
  journal = {Phys. Rev. A},
  volume = {81},
  issue = {4},
  pages = {041803(R)},
  numpages = {4},
  year = {2010},
  month = {Apr},
  publisher = {American Physical Society},
  doi = {10.1103/PhysRevA.81.041803},
  url = {https://link.aps.org/doi/10.1103/PhysRevA.81.041803}
}

@article{Zhu2025,
  title = {Autonomous quantum heat engine enabled by molecular optomechanics and hysteresis switching},
  author = {Zhu, Baiqiang and Meystre, Pierre and Zhang, Weiping and Zhang, Keye},
  journal = {Phys. Rev. A},
  volume = {112},
  issue = {4},
  pages = {043508},
  numpages = {11},
  year = {2025},
  month = {10},
  publisher = {American Physical Society},
  doi = {10.1103/v66j-6fv4},
  url = {https://link.aps.org/doi/10.1103/v66j-6fv4}
}

@article{Huang2024,
  title = {Collective quantum entanglement in molecular cavity optomechanics},
  author = {Huang, Jian and Lei, Dangyuan and Agarwal, Girish S. and Zhang, Zhedong},
  journal = {Phys. Rev. B},
  volume = {110},
  issue = {18},
  pages = {184306},
  numpages = {10},
  year = {2024},
  month = {11},
  publisher = {American Physical Society},
  doi = {10.1103/PhysRevB.110.184306},
  url = {https://link.aps.org/doi/10.1103/PhysRevB.110.184306}
}

@article{Kippenberg2005,
  title = {Analysis of Radiation-Pressure Induced Mechanical Oscillation of an Optical Microcavity},
  author = {Kippenberg, T. J. and Rokhsari, H. and Carmon, T. and Scherer, A. and Vahala, K. J.},
  journal = {Phys. Rev. Lett.},
  volume = {95},
  issue = {3},
  pages = {033901},
  numpages = {4},
  year = {2005},
  month = {Jul},
  publisher = {American Physical Society},
  doi = {10.1103/PhysRevLett.95.033901},
  url = {https://link.aps.org/doi/10.1103/PhysRevLett.95.033901}
}

@article{Montenegro2019,
  title = {Mechanical qubit-light entanglers in hybrid nonlinear qubit optomechanics},
  author = {Montenegro, Victor and Neto, G. D. de Moraes and Bose, Sougato},
  journal = {Phys. Rev. A},
  volume = {100},
  issue = {5},
  pages = {052310},
  numpages = {9},
  year = {2019},
  month = {Nov},
  publisher = {American Physical Society},
  doi = {10.1103/PhysRevA.100.052310},
  url = {https://link.aps.org/doi/10.1103/PhysRevA.100.052310}
}

@article{Molinares2025,
  author  = {H. Molinares and V. Eremeev},
  title   = {Drive-Loss Engineering and Quantum Discord Probing of Synchronized Optomechanical Squeezing},
  journal = {Mathematics},
  volume  = {13},
  pages   = {2171},
  year    = {2025},
  doi     = {10.3390/math13132171}
}

@article{Eremeev2025,
	author = {Eremeev, Vitalie and Molinares, Hugo and Correa, Luis A. and He, Bing},
	doi = {https://doi.org/10.1002/qute.202400497},
	journal = {Adv. Quant. Techn.},
	number = {3},
	pages = {2400497},
	title = {Simultaneous Photon and Phonon Lasing From Pumping Optomechanical Systems with a Two-Tone Field},
	volume = {8},
	year = {2025}}

@article{Kronwald2013,
  title = {Arbitrarily large steady-state bosonic squeezing via dissipation},
  author = {Kronwald, Andreas and Marquardt, Florian and Clerk, Aashish A.},
  journal = {Phys. Rev. A},
  volume = {88},
  issue = {6},
  pages = {063833},
  numpages = {10},
  year = {2013},
  month = {Dec},
  publisher = {American Physical Society},
  doi = {10.1103/PhysRevA.88.063833},
  url = {https://link.aps.org/doi/10.1103/PhysRevA.88.063833}
}

@article{Grudinin2010,
  title = {Phonon Laser Action in a Tunable Two-Level System},
  author = {Grudinin, Ivan S. and Lee, Hansuek and Painter, O. and Vahala, Kerry J.},
  journal = {Phys. Rev. Lett.},
  volume = {104},
  issue = {8},
  pages = {083901},
  numpages = {4},
  year = {2010},
  month = {Feb},
  publisher = {American Physical Society},
  doi = {10.1103/PhysRevLett.104.083901},
  url = {https://link.aps.org/doi/10.1103/PhysRevLett.104.083901}
}

@article{Marquardt2007,
  title = {Quantum Theory of Cavity-Assisted Sideband Cooling of Mechanical Motion},
  author = {Marquardt, Florian and Chen, Joe P. and Clerk, A. A. and Girvin, S. M.},
  journal = {Phys. Rev. Lett.},
  volume = {99},
  issue = {9},
  pages = {093902},
  numpages = {4},
  year = {2007},
  month = {Aug},
  publisher = {American Physical Society},
  doi = {10.1103/PhysRevLett.99.093902},
  url = {https://link.aps.org/doi/10.1103/PhysRevLett.99.093902}
}

@article{Kuzyk2017,
  title = {Controlling multimode optomechanical interactions via interference},
  author = {Kuzyk, Mark C. and Wang, Hailin},
  journal = {Phys. Rev. A},
  volume = {96},
  issue = {2},
  pages = {023860},
  numpages = {7},
  year = {2017},
  month = {Aug},
  publisher = {American Physical Society},
  doi = {10.1103/PhysRevA.96.023860},
  url = {https://link.aps.org/doi/10.1103/PhysRevA.96.023860}
}

@article{Xiao2026,
  title = {Staircase mechanical energy growth in optomechanical systems of intermediate mechanical frequencies},
  author = {Xiao, Yi and Wu, Yi and Zhan, Qi-Kai and Zhang, Jin Lian and He, Bing and Lin, Qing},
  journal = {Phys. Rev. A},
  volume = {114},
  issue = {4},
  pages = {043506},
  numpages = {11},
  year = {2026},
  month = {Oct},
  publisher = {American Physical Society},
  doi = {10.1103/dc8z-cycp},
  url = {https://link.aps.org/doi/10.1103/dc8z-cycp}
}

@article{PhysRevA.35.5288,
  title = {Routh-Hurwitz criterion in the examination of eigenvalues of a system of nonlinear ordinary differential equations},
  author = {DeJesus, Edmund X. and Kaufman, Charles},
  journal = {Phys. Rev. A},
  volume = {35},
  issue = {12},
  pages = {5288--5290},
  numpages = {0},
  year = {1987},
  month = {6},
  publisher = {American Physical Society},
  doi = {10.1103/PhysRevA.35.5288},
  url = {https://link.aps.org/doi/10.1103/PhysRevA.35.5288}
}

@article{Shlesinger2023,
author = {Ilan Shlesinger  and Jente Vandersmissen  and Eitan Oksenberg  and Ewold Verhagen  and A. Femius Koenderink},
title = {Hybrid cavity-antenna architecture for strong and tunable sideband-selective molecular {R}aman scattering enhancement},
journal = {Science Advances},
volume = {9},
number = {51},
pages = {eadj4637},
year = {2023},
doi = {10.1126/sciadv.adj4637},
URL = {https://www.science.org/doi/abs/10.1126/sciadv.adj4637}}

@article{acsnano.4c16535,
    author = {Jakob, Lukas A. and Juan-Delgado, Adri{\'{a}}n and Mueller, Niclas S. and Hu, Shu and Arul, Rakesh and Boto, Roberto A. and Esteban, Ruben and Aizpurua, Javier and Baumberg, Jeremy J.},
    title = {Optomechanical Pumping
of Collective Molecular Vibrations
in Plasmonic Nanocavities},
    journal = {ACS Nano},
    volume = {19},
    number = {11},
    pages = {10977-10988},
    year = {2025},
    month = {3},
    issn = {1936-0851},
    doi = {10.1021/acsnano.4c16535},
    url = {https://doi.org/10.1021/acsnano.4c16535}
}

@article{Yu2026,
  title = {Strong Molecule-Light Entanglement with Molecular Cavity Optomechanics},
  author = {Yu, Hong-Yun and Jiao, Ya-Feng and Wang, Jie and Li, Feng and Yin, Bin and Liu, Qi-Rui and Jiang, Tian and Jing, Hui and Wei, Ke},
  journal = {Phys. Rev. Lett.},
  volume = {136},
  issue = {1},
  pages = {013602},
  numpages = {7},
  year = {2026},
  month = {1},
  publisher = {American Physical Society},
  doi = {10.1103/4z8v-f6s5},
  url = {https://link.aps.org/doi/10.1103/4z8v-f6s5}
}

@article{Dezfouli2019,
    author = {Dezfouli, Mohsen Kamandar and Gordon, Reuven and Hughes, Stephen},
    title = {Molecular Optomechanics in the Anharmonic Cavity-{QED}
Regime Using Hybrid Metal--Dielectric Cavity Modes},
    journal = {ACS Photonics},
    volume = {6},
    number = {6},
    pages = {1400-1408},
    year = {2019},
    month = {5},
    issn = {2330-4022},
    doi = {10.1021/acsphotonics.8b01091},
    url = {https://doi.org/10.1021/acsphotonics.8b01091}
}

@article{Barreda2021,
author = {Angela I. Barreda and Mario Zapata-Herrera and Isabelle M. Palstra and Laura Mercad\'{e} and Javier Aizpurua and A. Femius Koenderink and Alejandro Mart\'{i}nez},
journal = {Photon. Res.},
number = {12},
pages = {2398--2419},
publisher = {Optica Publishing Group},
title = {Hybrid photonic-plasmonic cavities based on the nanoparticle-on-a-mirror configuration},
volume = {9},
month = {12},
year = {2021},
url = {https://opg.optica.org/prj/abstract.cfm?URI=prj-9-12-2398},
doi = {10.1364/PRJ.433761},
}

@article{HuangLeiZhang2025,
    author = {Huang, Jian and Lei, Dangyuan and Zhang, Zhedong},
    doi = {10.1021/acs.nanolett.5c04075},
    journal = {Nano Letters},
    number = {46},
    pages = {16372--16378},
    title = {Generating Coherent {R}aman Scattering Using a Molecular Optomechanical Cavity},
    volume = {25},
    year = {2025}}

@article{PatraKafleHabteyes2023,
    author = {Patra, Bisweswar and Kafle, Bijesh and Habteyes, Terefe G.},
    doi = {10.1021/acs.nanolett.3c01035},
    journal = {Nano Letters},
    number = {11},
    pages = {5108--5115},
    title = {Molecular Optomechanics Induced Hybrid Properties in Soft Materials Filled Plasmonic Nanocavities},
    volume = {23},
    year = {2023}}

@article{Kinslow2026,
    author = {Kinslow, Joshua C. and Humphries, Ben S. and Green, Dale and Borgh, Magnus O. and Jones, Garth A.},
    doi = {10.1021/acs.jpclett.6c02396},
    journal = {The Journal of Physical Chemistry Letters},
    number = {41},
    pages = {11723--11730},
    title = {Limits of Memory-Assisted Coherence in Cavity-Embedded Molecules},
    volume = {17},
    year = {2026}}

@article{Casteels2016,
  title = {Power laws in the dynamic hysteresis of quantum nonlinear photonic resonators},
  author = {Casteels, W. and Storme, F. and Le Boit{\'e}, A. and Ciuti, C.},
  journal = {Phys. Rev. A},
  volume = {93},
  issue = {3},
  pages = {033824},
  numpages = {9},
  year = {2016},
  month = {Mar},
  publisher = {American Physical Society},
  doi = {10.1103/PhysRevA.93.033824},
  url = {https://link.aps.org/doi/10.1103/PhysRevA.93.033824}
}

\end{document}